\documentclass[letterpaper]{article}
\usepackage{aaai}
\usepackage{times}
\usepackage{helvet}
\usepackage{courier}
\usepackage{graphicx}
\usepackage{subfigure}
\usepackage{textcomp}
\usepackage{ifpdf}

\usepackage[hyphens]{url}
\usepackage{color}
\graphicspath{ {./figures/}{./} }
\title{Coding Together at Scale:\\GitHub as a Collaborative Social Network}

\author{
Antonio Lima, Luca Rossi and Mirco Musolesi \\
School of Computer Science\\
University of Birmingham, UK\\
\{a.lima,l.rossi,m.musolesi\}@cs.bham.ac.uk
}

\begin{document}
\maketitle

\begin{abstract}
\begin{quote}
GitHub is the most popular repository for open source code~\cite{finley_github_2011}. It has more than 3.5 million users, as the company declared in April 2013, and more than 10 million repositories, as of December 2013. It has a publicly accessible API and, since March 2012, it also publishes a stream of all the events occurring on public projects. Interactions among GitHub users are of a complex nature and take place in different forms. Developers create and fork repositories, push code, approve code pushed by others, bookmark their favorite projects and follow other developers to keep track of their activities.

In this paper we present a characterization of GitHub, as both a social network and a collaborative platform. To the best of our knowledge, this is the first quantitative study about the interactions happening on GitHub. We analyze the logs from the service over 18 months (between March 11, 2012 and September 11, 2013), describing 183.54 million events and we obtain information about 2.19 million users and 5.68 million repositories, both growing linearly in time. We show that the distributions of the number of contributors per project, watchers per project and followers per user show a power-law-like shape. We analyze social ties and repository-mediated collaboration patterns, and we observe a remarkably low level of reciprocity of the social connections. We also measure the activity of each user in terms of authored events and we observe that very active users do not necessarily have a large number of followers. Finally, we provide a geographic characterization of the centers of activity and we investigate how distance influences collaboration.
\end{quote}

\end{abstract}

% flatex input: [intro.tex]
\section{Introduction}

In recent years, GitHub\footnote{https://github.com/}, a hosting platform for software projects, has gained much popularity among a large number of software developers around the world. This platform offers version control hosting, as other platforms have done in the past (e.g., SourceForge\footnote{https://sourceforge.net/}, Assembla\footnote{https://www.assembla.com/}, BitBucket\footnote{https://bitbucket.org/}). However, this service has much emphasis on its social features, as summarized in its motto ``GitHub: social coding''. In fact, GitHub is not simply offering a code hosting service, like its competitors had been doing for a long time, but also an easy-to-use and cheap (or even free in its basic version) online tool for collaborative software development and many features supporting the community of developers. For all these reasons, GitHub has successfully lowered the barrier to collaboration in open source. The importance of this collaboration platform seems to be increasing, as its founder has plans to extend the use cases beyond software development~\cite{techcrunch_github_2013}.
At the same time, most of the data concerning collaboration on public\footnote{GitHub also offers fee-based private repositories. Since it is not possible to access any information about private repositories, our analysis will focus on public repositories.} software repositories can be accessed and analyzed. This represents a unique opportunity to study aspects of human behavior related to collaboration at scale.

GitHub is based on the Git revision control system\footnote{http://git-scm.com/}. In GitHub a user can \textit{create} code repositories and \textit{push} code to them. Every repository has a list of \textit{collaborators}; they can make changes to the content of the repository and they review the contributions that are submitted to the repository, accepting or discarding them. In this sense, they are not the only people collaborating on the project. In fact, every person that wishes to contribute to a project, without being a collaborator, can \textit{fork} it\footnote{In the open-source context, the term \textit{forking} had a negative connotation in the past, i.e., it has often been used to refer to groups of developers separating for different views on a project and splitting their forces on independent projects. Instead, in GitHub, forking is a normal part of the process of contributing to a project.
%Indeed, GitHub encourages to use the ``Fork \& Pull'' Collaborative Model, rather than the ``Shared Repository Model''. More information can be found at: https://help.github.com/articles/using-pull-requests
}.
This action makes a duplicate of the repository, allowing developers to work independently, \textit{committing} changes only to their own fork. When developers complete a certain task (e.g., a new feature or a bug fix), they can send the changes to the original repository, through a so-called \textit{pull request}. Then, a collaborator of the original repository reviews the changes contained in the pull request and decides whether to accept it in the original repository (in the Git jargon, \textit{merge} it to the parent repository),  or refuse it, optionally motivating his or her choice. Once the new code is accepted in the original repository, its author becomes one of the \textit{contributors} of the project.
%The whole process usually involves the creation of ``branches''
%but we will not go into that to keep the description simple, as we do not need the concept of branches for the analysis we intend to make.
%but in order to simplify the discussion in the remainder of this work, we will not take into consideration that concept.\mm{Since we are not discussing branches, we might think to introduce them here in the first place perhaps?}
%
In addition to that, GitHub users can \textit{follow} other users, to be notified of their actions. The website is not used only for collaboration, but also as a resource to find quality software. Users can \textit{star} interesting repositories that they want to bookmark for later reference. Other features are also available (e.g., issue tracking, downloads, gists, and so on) but we will not consider them in this work.

In this paper, to the best of our knowledge, we present the first in-depth quantitative analysis of GitHub, as a unique example of large-scale real-world collaboration platform mainly used for software projects. The contributions we make in this paper can be summarized as follows:
\begin{itemize}
\item We conduct basic structural analyses and we show that the distributions of the number of contributors and watchers per project and followers per user show a power-law-like shape.
\item We analyze social ties and repository-mediated collaboration patterns. We find a very low reciprocity of the social ties, which is remarkably different from the findings of studies of other types of social networks.
\item We study the depth and width of the trees corresponding to forked repositories and we observe that in GitHub collaboration on forks seems to happen on a limited number of key projects.
\item We investigate the correlation between the activity of users and their popularity in the network and we observe that very active users do not necessarily have a large number of followers.
\item We provide a geographic characterization of activities and collaborations. We find that users tend to interact with people they are close to and that repositories with a low number of collaborators tend to have them concentrated around a few specific geographic locations, rather than scattered around the world. Finally, we observe a similarity between the geographical distributions of following and contributions ties.
\end{itemize}

The paper is organized as follows. We first discuss related work in this area. Then, we present the data collection methodology and we describe the characteristics of the dataset. We study the networks representing the interactions between entities, extracted from the dataset. We conclude the article by discussing our key findings and outlining our future work.

%The paper is organized as follows. In Section~\ref{sec:related} we discuss related work. In Section~\ref{sec:dataset} we present the data collection methodology, and we describe the characteristics of the dataset. In Section~\ref{sec:results}
%we study the networks representing the interactions between entities, generated from the dataset. Finally, in Section~\ref{sec:conclusion} we conclude by discussing our findings.
% flatex input end: [intro.tex]

%
% flatex input: [related.tex]
\section{Related Work}
\label{sec:related}

Several researchers from different communities have been interested in analyzing behavior on websites and online tools that enable large-scale collaboration, most notably Wikipedia. Indeed, a large body of research has focused on understanding how people coordinate their collaboration efforts in the constant update and expansion of the crowdsourced online encyclopedia through a variety of methodologies (see for example~\cite{kittur_power_2007,vuong_ranking_2008}). A relevant approach in relation to the topic of this work is the network analysis of the collaboration structure in Wikipedia presented in~\cite{brandes2009network}.
%
% In previous literature, the collaboration of developers over projects has been
% analysed mainly from two point of views: the analysis of the social network
% emerging from the interaction and communication within the project, 
% and the analysis of contributions to the project. A large parte of
% previous work has taken into consideration the collaboration on single software
% projects, with the intent of giving measures and visualizations that help understand
% how contribution happens on the repository. Some recent works have analysed the
% data generated by project hosting websites, such as SourceForge, Google Code
% and, lastly, GitHub.
%
More in general, open-source projects have been the subject of
several studies specifically aimed at uncovering the social structure that
emerges from the interactions between developers~\cite{valverde2007self,bird2008latent} and
at analyzing the individual contributions to specific
projects~\cite{hindle_what_2008}.% WIth the rise of 

%Kwak et al.~\cite{kwak2010twitter}

Recently, given its increasing popularity, there has been a surge in interest in GitHub and its underlying social dynamics. Some projects are currently undergoing with the specific aim of providing easy-to-use and efficient tools for accessing data from GitHub, especially in real-time. For example, \cite{gousios2012ghtorrent} discusses a system to gather streams and data from GitHub in a scalable fashion to overcome the limitations imposed by the GitHub API, specifically directed at researchers.
 
In~\cite{dabbish2012social} an in-depth qualitative user study is conducted on a small group of GitHub users, aimed at understanding the motivations that are the basis of online collaboration and the consequences of using a transparent large-scale tool on the practice of software development. They find that people use GitHub for several reasons: to learn how to code better, to follow popular developers, to find new interesting projects, and to promote themselves and their work. They also find that actions in GitHub, such as following a user, committing changes and accepting/rejecting code, are heavily influenced by specific characteristics of the interactions happening in the system. Some other studies have investigated the geographical distribution of collaborations~\cite{heller2011visualizing,shrestha2013visualizing}. An example is that of Heller et al.~\cite{heller2011visualizing}, who use visualization techniques to identify patterns of collaboration, including geographic characteristics of the interactions between cities and influence among them.

With respect to this body of work, to the best of our knowledge, our paper presents the first systematic quantitative analysis of the interactions in GitHub. 
% from a network modeling perspective\al{This whole sentence must be rephrased. We do not make any modeling, hence I would avoid to use this word.}, also analyzing the impact of geography on the patterns of collaboration, by means of extensive measurements on a large-scale longitudinal dataset.
We believe that our quantitative methodology complements the existing qualitative work by providing insights about global patterns of interactions that are not possible to obtain by means of small-scale and interview-based studies.
% flatex input end: [related.tex]

%
% flatex input: [dataset.tex]
\section{Description of the Dataset}
\label{sec:dataset}
\begin{figure}[t]
\includegraphics[width=\linewidth]{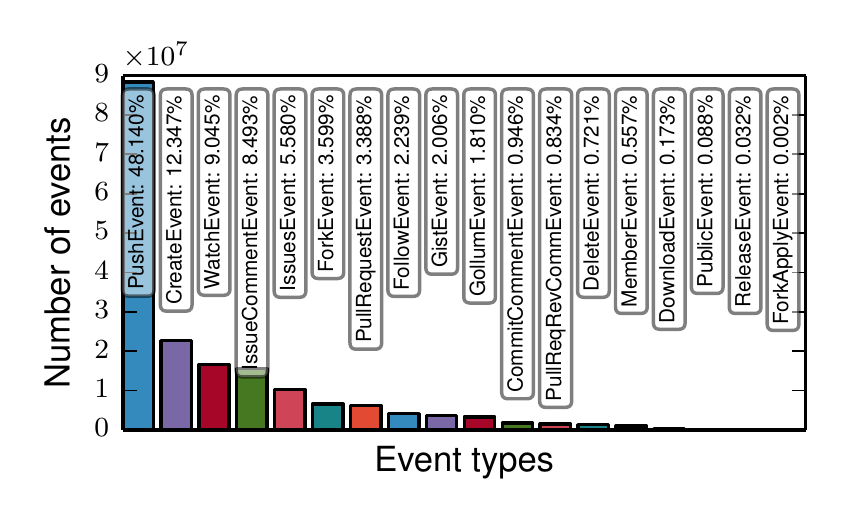}
    \caption{Number of events detected in the GitHub stream.}
\label{fig:events-histogram}
\end{figure}

The full list of public events that have happened on GitHub is available on the GitHub Archive website\footnote{\url{http://www.githubarchive.org}}. In this paper, we analyze events that happened on GitHub over a period of 18 months, between March 11, 2012 and September 11, 2013, retrieved from that archive. Our dataset includes various types of events performed by users on public repositories or following events between users (i.e., when a user starts following another user). The total number of retrieved events is 183,540,210 and they fall into 18 categories\footnote{\url{http://developer.github.com/v3/activity/events/types/}}. Each event, regardless of its kind, usually includes some metadata about the entities involved (e.g., the profile information of a user, his or her number of followers, the language of a repository, etc.).
%With each event we update the metadata we hold about users and repositories.\mm{This sentence is not clear - it is term "update" that should be clarified. Maybe something like "From the events, we extract the interactions between the users and the activities on the repositories over time." }
Fig.~\ref{fig:events-histogram} shows how events are distributed among the various categories. %Based on this dataset, we are able to reconstruct the variations in time of various networks, that we will introduce in the next section.
%\mm{I suggest to remove this since we have not defined the networks at this point of the paper}
%
One outlier user under the name of \texttt{Try-Git} shows an uncommonly high number of collaborations. As it is a learning tool that pushes code automatically to other users' repositories, we discarded it from the dataset.
%It is also worth noting that in our analysis we do not give particular meaning to operations performed on different branches of the same repository and we include all of them in the analysis.

% GEOCODING
In order to explore the geographic features of users, we investigate the location information that can be found in the user profiles. In our dataset, 345,625 users have a non-empty location field. As the field is optional, there is little incentive to fill it with fake information. Therefore, we can reasonably assume that most of the non-empty entries are truthful. In order to convert the text field to an unambiguous location, we use the MapQuest Open Geocoding API\footnote{Data:~\textcopyright{~OpenStreetMap} contributors, available under the Open Database License. Geocoding: courtesy of MapQuest (\url{http://www.mapquest.com}).}. We evaluate the validity of the geocoder by considering a sample of 1,000 users in the population of users with non-empty location field and assessing the fraction of correctly geocoded elements by manually labeling them. We find that 106 elements are incorrectly geocoded. From the analysis of this sample, therefore, we can say that the geocoder fails to correctly convert to coordinates in $10.6 \pm 1.91 \%$ cases of the original population, with $95\%$ confidence level. Incorrectly geocoded entries in the sample fail mostly for the following reasons: because they describe multiple locations (for example \texttt{"London and Nottingham"}), because they have no geographic meaning (e.g., \texttt{"localhost", "emacs"}) because they are ambiguous (e.g., \texttt{"San Jose"}, rather than \texttt{"San Jose, CA"}).

%\item
%\al{This has either to be removed or updated, as the data is old now}. As a second data source (SET2), we consider the information of popular repositories. We also crawled data from GitHub API, choosing the 50 repositories with the highest number of forks and 50 repositories with the highest amount of
%stars, amounting to 67 unique repositories. For each repository we
%downloaded information about the contributors and the stargazers. In the Github
%terminology, contributors are people who have committed code to the project;
%stargazers are users who starred the project, and therefore are ``following''
%the project. For each contributor and stargazers, we geocoded their user
%location field and we crawled their ingoing and outgoing social links.

%\subsection{Collaboration inequality}
%
% In most open source projects a small group of developers accounts for most of the contributions and usually leads the project. The Gini coefficient, a measure of statistical dispersion commonly used in economy to measure inequality of distribution. Here we use it to characterize how unevenly the workload is shared across contributors.
% \begin{equation}
% G = \frac{1}{2E\bar{x}}\sum_i{\sum_j{|x_i - x_j|}}
% \end{equation}
% where $E$ is the total number of contributors, $\bar{x} = \sum_i x_i/E$ is the average number of contributions per user and $i$ and $j$ represent contributors labels.

% flatex input end: [dataset.tex]

%
% flatex input: [results.tex]
%\section{Results}
%\label{sec:results}
%
\begin{figure}[t]
\includegraphics[width=\linewidth]{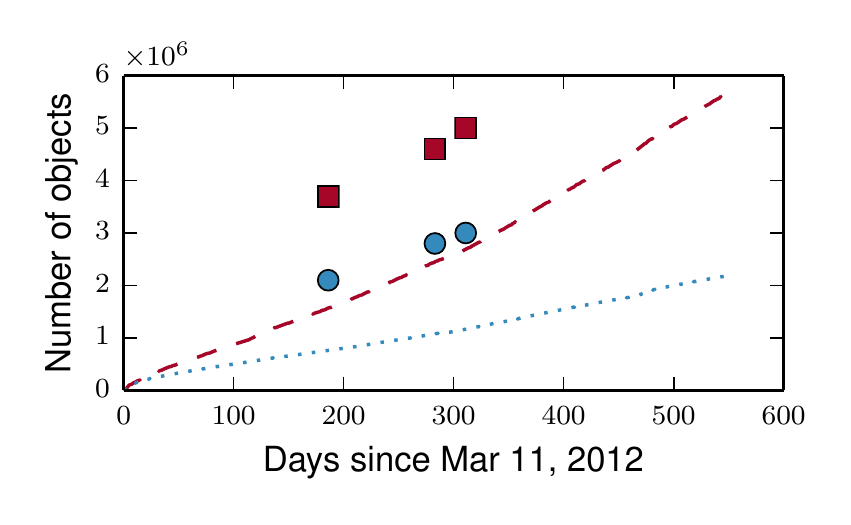}
    \caption{Number of unique repositories and unique users detected from the stream since March 11, 2012. The dashed and dotted blue lines show the number of repositories and the number of users detected from the event stream, respectively. The three squares and circles indicate the number of repositories and users in three specific dates as advertised by GitHub itself on its website.}
\label{fig:evolution}
\end{figure}
%
%{Evolution of the Dataset}
%
%\mm{I do not know if we want to put here a short paragraph saying why we perform the temporal analysis, a sort of motivation/justification of this analysis}
%\subsection{Evolution of the Service}
%As described in the previous section, we collect our dataset from the event stream. 
%Therefore, before going further in the analysis, it is appropriate to investigate as the dataset changes as the stream taken into consideration grows longer in time. 
%In this section, we present an analysis of the temporal analysis of the dataset, which provides interesting insights about the dynamics of growth of GitHub itself.\mm{modified}
%
It is important to be aware that this data source suffers a time bias, since the archive does not include events happened before March 2011. %We will make considerations about this bias in the next section, while describing the temporal evolution of the networks.
In Fig.~\ref{fig:evolution} we show the number of unique users and public repositories seen in the event stream since March 11, 2012. 
As previously discussed, we are able to retrieve metadata when entities are involved in an \textit{event}. In other words, we do not have information about dormant entities that were created before March 11, 2012 and do no longer generate \textit{any} event during the subsequent 18 months (e.g., an inactive user, an abandoned repository). We are also not able to extract pre-existing following relations from the stream. After a short transitory period, which is present because of the temporal bias of our data collection process based on events, both curves show linear growth with different coefficients, with the ratio describing the number of repositories over the number of users reaching a steady value of approximately 2.59. The figure also reports the number of repositories and users (indicated using squares and circles, respectively) publicly declared by GitHub.
In our dataset, we observe a lower number of users and repositories for two reasons. Firstly, the official numbers include \textit{all} the users and repositories created since the launch of the website in 2008, whereas our dataset contains only the \textit{active} users and repositories in the period taken into consideration. Secondly, the official statistics probably include private repositories, that do not produce events on the public timeline we are able to access. For these reasons, we can conclude that a large number of users do not actively use the website (i.e., do not generate events) or they act exclusively on private repositories. These figures also suggest that a large number of repositories are either abandoned or private.
% https://english.stackexchange.com/questions/5378/a-number-of-questions-has-been-or-have-been-asked
%[WEEK, MONTH PERIODICITY]

%[What is the growth of CONSTANTLY ACTIVE users/repositories? Can we remove dormient entites from the stream (for example, a user that is active for a month, then leaves)]

%[Another point we need to make somehow is that interaction happens really on a very minor set of popular repositories. How many are them? How can we define the activity?]

\section{Structural Analysis}

In this section we define, extract and analyze several networks, generated from the event stream, which describe interactions between users and repositories.
\begin{itemize}
\item We represent users' following relations by means of a directed graph $G_F$, which we call \textit{followers graph}. We are able to reconstruct this network by looking at \textit{follow events} in the stream.
\item We represent the collaborations of users on repositories as a bipartite graph $G_C$, the \textit{collaborators graph}, where repository nodes are connected to their collaborators nodes. We are able to infer this network by extracting from \textit{push events} information about who uses write permission and on which repositories. We refer to $G_C^{\perp}$, the \textit{projected collaborators graph}, as to the graph obtained by projecting the collaborators graph onto the set of users. In this projected graph users who collaborate in at least one repository are connected to each other.
\item We represent users assigning a star to a repository as a bipartite graph $G_S$, the \textit{stargazers graph}. This network can be generated using the information found in \textit{watch events}.
\item Finally, we build the \textit{contributors graph} $G_N$ by analyzing the content of every \textit{push event}, which includes authorship information of the pushed commits.
\end{itemize} 
For our static analyses we consider these networks as they appear on the final day of the time window we take into consideration.
%\al{add small figure perhaps.}

%Starting from GitHub events, we can build several networks, which represent various forms of interaction between entities. In this section we will describe how we built each network and we will analyze it. 
%In this section we investigate the structural properties of the networks that can be built from . More specifically, we analyze both the followers graph and the graph obtained by projecting the bipartite contributors graph onto the set of nodes representing the users. In some sense, the followers graph can be seen as an explicit representation of the social network underlying GitHub. On the other hand, some sort of social structure is also implicitly defined by the connections induced by the collaborations of the users on the repositories. Thus, in this section we study these two networks and we try to establish to what extent they share similar characteristics.

\subsection{Followers and Collaborators Networks}
\begin{figure}[t]
\includegraphics[width=\linewidth]{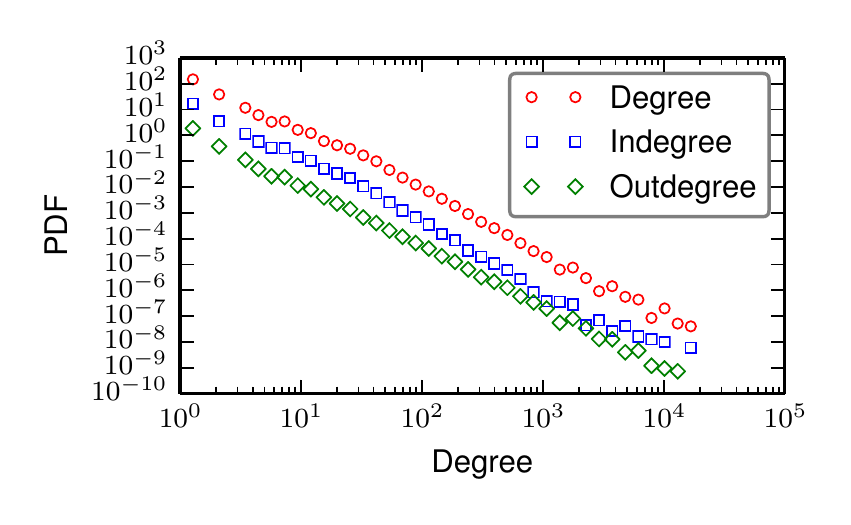}
    \caption{Distribution of degree, in-degree and out-degree of the social graph. The distributions were shifted along the y-axis to put in evidence their structure. The three distributions exhibit a power-law scaling behavior, with different exponents, for values in the range from 20 to 1000.}
\label{fig:socialgraph-distributions}
\end{figure}
\begin{figure}[t]
\includegraphics[width=\linewidth]{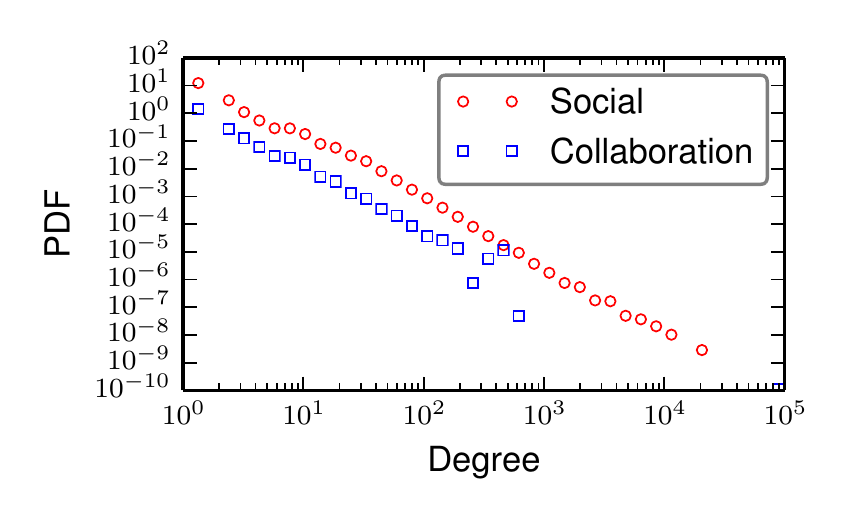}
    \caption{Distribution of the number of followers per users (red) and the number of total collaborators per user, (blue) which corresponds to the degree distribution of the users projection of the collaborators bipartite graph.}
\label{fig:collaboration-distribution}
\end{figure}

As previously explained, a user follows other users in order to be regularly updated about events regarding them (e.g., forks, created repositories, starred repositories, and so on). The followers graph $G_F$ we obtain has a total of 671,751 nodes and 2,027,564 edges, with a resulting graph density of 4.4932e-06 and an average degree of 3.019. The low graph density and average degree indicate that on GitHub the follow action is associated with a high cost, as following many developers results in receiving many notifications from them. This result also reflects the fact that following links in GitHub do not play the same important role they have in other social networks, such as Facebook or Twitter.
%As we will discuss later in detail, interactions between GitHub users are mostly mediated by repositories.\al{Check that we do discuss this.}\lr{I don't think we do, nor we can do it. we should remove the claim if we don't have more data (how many pairs of users have both a following edge and a collaboration edge? how about the other way around?)}

%\begin{figure*}[t]
%\centering
%\subfigure[Write me]{\includegraphics[width=.3\linewidth]{img1}}
%\subfigure[Write me]{\includegraphics[width=.3\linewidth]{img2}}
%\subfigure[Write me]{\includegraphics[width=.3\linewidth]{img3}}
%\caption{Write me.}
%\label{name_me}
%\end{figure}

\begin{figure*}[t]
\centering
\subfigure[Followers]{\includegraphics[width=\columnwidth]{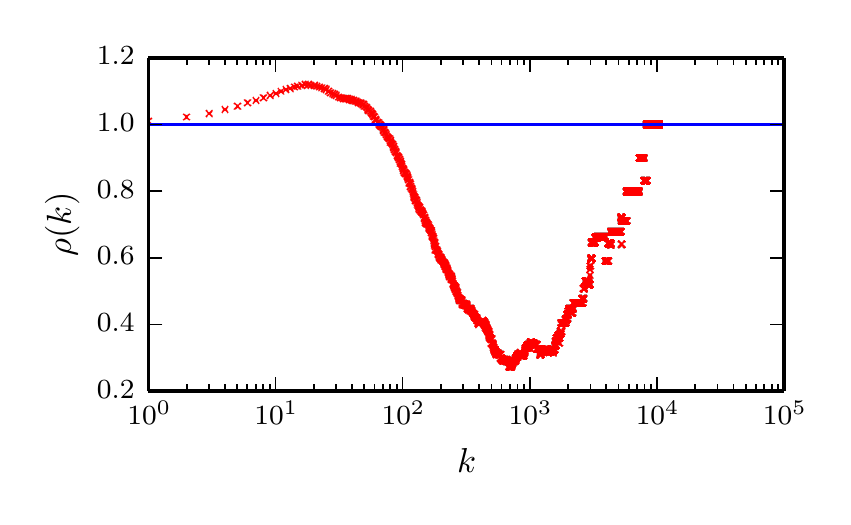}\label{fig:social-rich}}
\subfigure[Collaborators]{\includegraphics[width=\columnwidth]{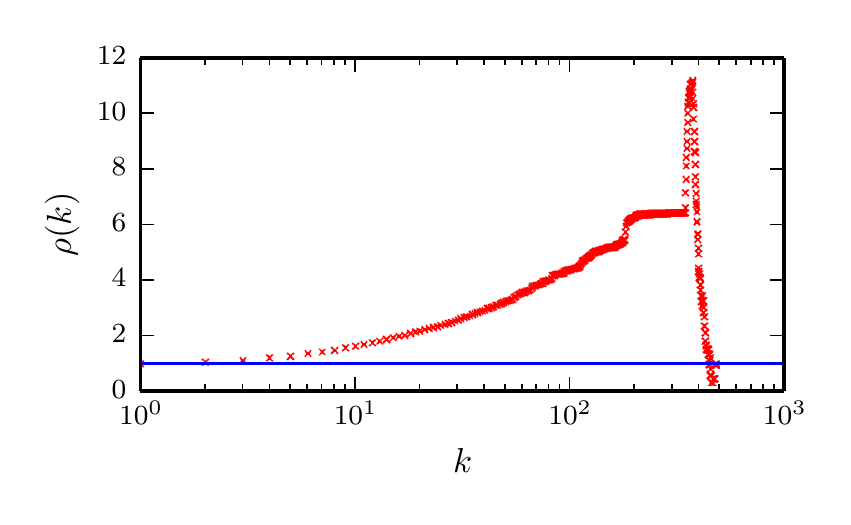}\label{fig:collaborators-rich}}
\caption{The normalized rich-club coefficient $\rho(k)$ as a function of the node degree. The blue horizontal line represents the value of $\rho(k)$ on a maximally random graph. Values of $\rho(k)$ above (below) 1 correspond to the presence (absence) of the rich-club phenomenon with respect to the random case. The two networks show remarkably different rich-club behaviors, due to their distinct nature.}
\end{figure*}
Fig.~\ref{fig:socialgraph-distributions} shows the distributions of the in-degree, out-degree and total degree of the users in $G_F$. All the three distributions show a power-law scaling behavior, characterized by different regimes.
%The although, with different regimes \al{actually the outdegree has only a regime} 	\lr{actually I barely see the different regimes of the others two, to be honest. should we avoid entirely to mention the regimes thing?}. 
We also note that for degrees smaller than $k \approx 20$, in all the three cases the scaling relation is not satisfied. Interestingly, we also find that the degree distributions of $G_F$ and of $G_C^{\perp}$ follow the same power-law regime, as shown in Fig.~\ref{fig:collaboration-distribution}. However, the node degree in the followers graph grows considerably larger than in $G_C^{\perp}$.

The followers network is also characterized by low reciprocity: only $9.6\%$ of the pairs of users have a reciprocal relation between them, while the remaining $90.4\%$ are one-way. Other studies on social networks reported considerably higher levels of reciprocity, such as $22.1\%$ for Twitter~\cite{kwak2010twitter}, $68\%$ for Flickr~\cite{cha2009measurement} and $84\%$ for Yahoo! 360~\cite{kumar2010structure}. The consistently lower reciprocity in Twitter is partially motivated by the presence of a few popular programmers, the so-called ``rockstar programmers'', who exhibit high in-degrees and low out-degrees. However, we believe the profoundly different nature of GitHub, compared to other social networks, might also play a role in this. In fact, social networks are mostly used for leisure and they thrive on distractions coming from noisy timelines; on the contrary, the productivity of GitHub developers might be critically disrupted by non-relevant notifications, which are hence kept to a minimum. In other words, establishing links has high cost in GitHub, as people do not ``follow-back'' unless they are professionally interested in the activity of their followers.

In order to uncover the presence of node degree correlations, we first measure the degree assortativity. We say that a network shows an assortative mixing~\cite{newman2002assortative} if nodes with a large number of links tend to share edges with high degree nodes. Similarly, if nodes with a small number of links tend to share edges with low degree nodes we say that a network shows a disassortative mixing. In our case we find a value of -0.0386, which suggests a tendency to a disassortative mixing of users. 
We also evaluate the rich-club coefficient $\phi$~\cite{zhou2004rich}, which measures the tendency of high degree nodes to form tightly interconnected communities.
%, and it is a key property for the formation of communities in social networks. 
% \mm{I would say that it depends on the same type of networks - since we should add more details, I suggest to remove it.
Although apparently similar to the concept of assortative mixing, the rich-club phenomenon is not necessarily associated with the latter, as one can define a disassortative network that still shows evidence of a rich-club phenomenon. Let $E_k$ denote the number of edges among the $N_k$ nodes having a degree higher than $k$. The rich-club coefficient $\phi(k)$ is defined as follows:

\begin{equation}
\phi(k)=\frac{2 E_k}{N_k (N_k - 1)}
\end{equation}

It represents the fraction of edges connecting nodes in $N_k$ out of the maximum possible amount they can share, i.e., $\frac{N_k (N_k -1 )}{2}$. More specifically, here we use the normalized rich-club coefficient proposed by~\cite{colizza2006detecting}, where the normalization is introduced to account for the fact that high degree nodes have a higher probability of sharing edges than low degree ones. Fig.~\ref{fig:social-rich} shows the rich-club index of the followers graph, for increasing degree $k$. We use the definition of the rich-club index for $G_F$ considering it as an undirected graph.
%Note that we compute the rich-club index on a symmetrized version of $G_F$, although recently this index was also extended to directed graphs~\cite{smilkov2010rich}. 
Interestingly, we see that low degree nodes show a less accentuated rich-club phenomenon, while high degree nodes do not. In other words, the plot indicates that hubs, i.e., popular developers, tend to share links with lower degree nodes rather than being tightly interconnected among them.

Compared to the followers graph, the collaborators graph $G_C^{\perp}$ also shows disassortative mixing of the nodes, with a value of $-0.0518$. However, the characteristics of the rich-club phenomenon are remarkably different. Fig.~\ref{fig:collaborators-rich} shows the rich-club index of $G_C^{\perp}$, for increasing values $k$ of the degree. We observe that up to $k\approx 30$ the nodes show a strong rich-club phenomenon, with a pronounced increase followed by a sudden drop around $k\approx 40$. This effect is amplified by the projection operation itself, as each group of collaborators forms a clique in $G_C^{\perp}$.  

We also measure the clustering coefficient~\cite{watts1998collective} of $G_C^{\perp}$ and we compare it with that of the followers graph. Again, we expect the average clustering coefficient of the network to be high due to the way in which $G_C^{\perp}$ is constructed. Indeed, we find a value of 0.395 for $G_C^{\perp}$ and of 0.012 for $G_F$. Note, however, that this implies that users contributing to the same repositories do not necessarily follow each other, as in that case we would expect the average clustering coefficients of the two networks to be similar. Once again, this underlines the fact that the social interactions captured by the two structures are rather different.

%\subsection{Further Comments}
%Although we showed that there is no clear correlation between the act of following and contributing, in that people who contribute to the same repository do not seem to follow each other, one can still ask if there exist a relation between the number of followers of a user and its contributions to GitHub. 
We now investigate the relation between the number of followers of a user and his/her contributions to GitHub.We would expect popular users in terms of contributions to be followed by a higher number of people. In order to evaluate this, we measure the Spearman correlation coefficient~\cite{lehmann2006nonparametrics} between the number of followers and the number of contributions per user, and we find a value of 0.2568, with p-value $<0.01$, indicating the lack of a clear correlation between the two dimensions. This result is unexpected, as it would seem reasonable to assume for \emph{active} users, i.e., users that contribute to a large number of repositories, to be more popular in terms of followers.

\subsection{Interactions on Repositories}

\begin{figure}[t]
\center
\includegraphics[width=\linewidth]{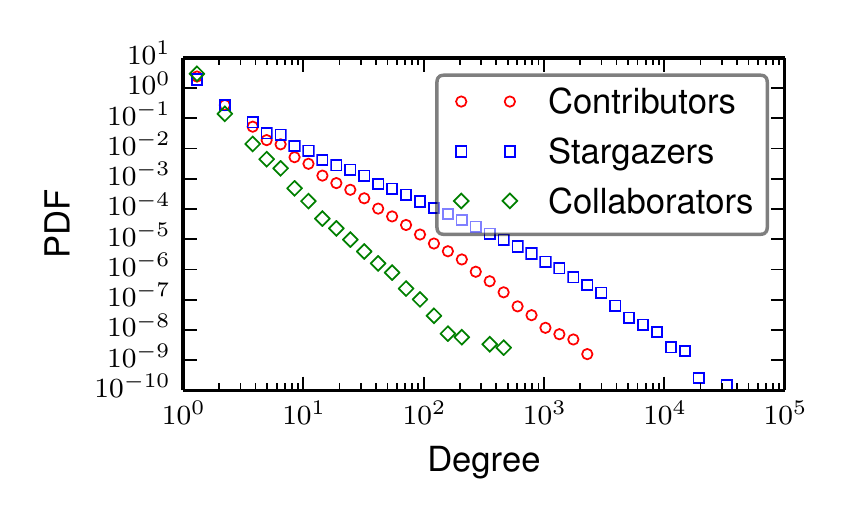}
\caption{Distribution of the number of contributors, collaborators and stargazers per repository. The contributors and stargazers distributions are best fitted by a power law distribution with exponential cut-off and exponent $\alpha$ equal to $2.34$ and $1.77$, respectively. The collaborators distribution is consistent with a power law with $\alpha = 3.39$. The distributions parameters are estimated using the approach of~\cite{clauset2009power}.}
\label{fig:contributors-stargazers}
\end{figure}

Despite the large number of repositories hosted at GitHub, developers work only on a consistently smaller fraction of them. Only 62.90\% of the total number of repositories we obtain information for experience at least one code commit during the 18 months taken into consideration. Only 74.22\% of these repositories have at least two contributors, meaning that one active repository out of four is exclusively authored by a single individual. This might happen for a variety of reasons: the project might not look promising to other users or the owners of the repository might reject contributions. This fraction includes activity both from one-time and habitual collaborators. Commonly, long-term contributors are turned into collaborators, so that they can help developing big projects. However, this kind of collaboration is quite rare, as only 9.61\% of the repository has at least 2 of them. This is not surprising: collaborators need to be trusted individuals who have full understanding of the project goals and structure, as they have write access on the repository and they determine which contributions should be accepted. Fig.~\ref{fig:contributors-stargazers} reports the distribution of number of contributors, stargazers and collaborators per repository.

\subsection{Forking and Repository Tree Structure}

The fork operation is intended to let users actively contribute to a project. This action produces a copy of the parent repository and essentially generates a simple tree structure. Further forks on the leaves of the tree increase its depth, while forking an internal node results in an increased width of the set of its children.  We interpret the repository tree as a directed acyclic graph, where the fork operation generates a directed edge from the parent repository to its child. In the following we refer to the \emph{depth} of the tree as the longest path from the root to its leaves, and to its \emph{width} as the maximum number of children over the internal nodes or 1 if the root has no children.

For a few repositories the maximum depth goes up to 12. However, these few structures are hardly the result of collaboration, in our opinion. In fact, user accounts involved in their creation do not exist anymore. For this reason, we suppose these accounts have been removed due to abnormal or suspicious activity. We also find that the average depth is 3.0695, but the mode is 0, indicating that the majority of repositories has a low number of contributions. The width, on the other hand, goes up to 10,256, which is normal considering that many people fork to contribute to popular packages, such as \texttt{mxcl/homebrew}. Top repositories include \texttt{heroku/node-js-sample}, \texttt{YOU-LOST/THE-GAME} (apparently, a ludic non-software repository) and \texttt{facebook-tornado}. The overall average width is very low (1.0653), showing that just a few popular repositories get forked, while the vast majority of them ($93.91\%$) have a width of just 1. This, together with the observation that the majority of the repositories has depth equal to $0$ and width equal to $1$, seems to suggest that forks on GitHub happen on a limited number of key projects.

\section{Activity, Social Presence and Indirect Rewards}
\label{rewards}

\begin{figure*}[t]
\centering
\subfigure[Events vs Followers]{\includegraphics[width=.45\linewidth]{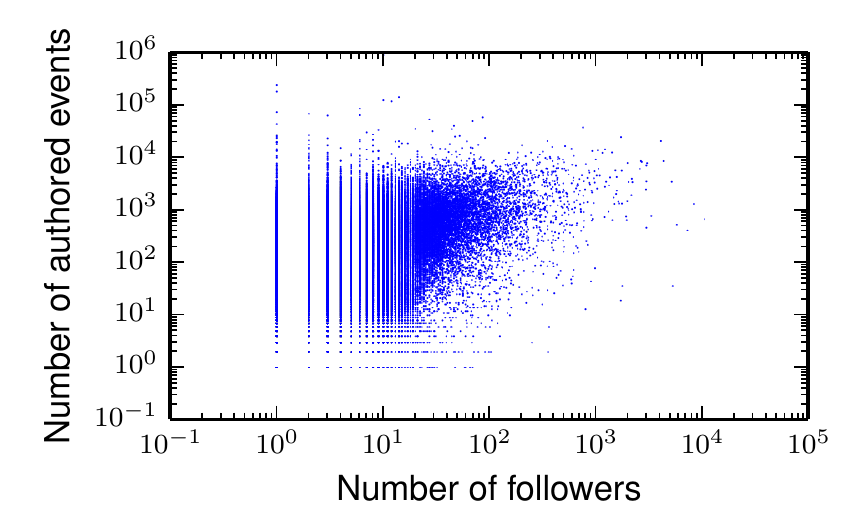}\label{fig:events-vs-followers}}
\subfigure[Events vs Collaborators]{\includegraphics[width=.45\linewidth]{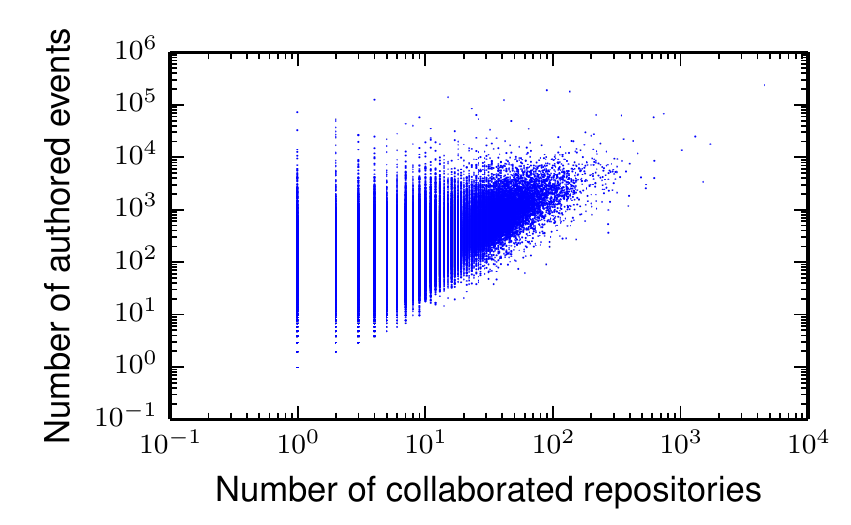}\label{fig:events-vs-collaborators}}
\subfigure[Events vs Followed users]{\includegraphics[width=.45\linewidth]{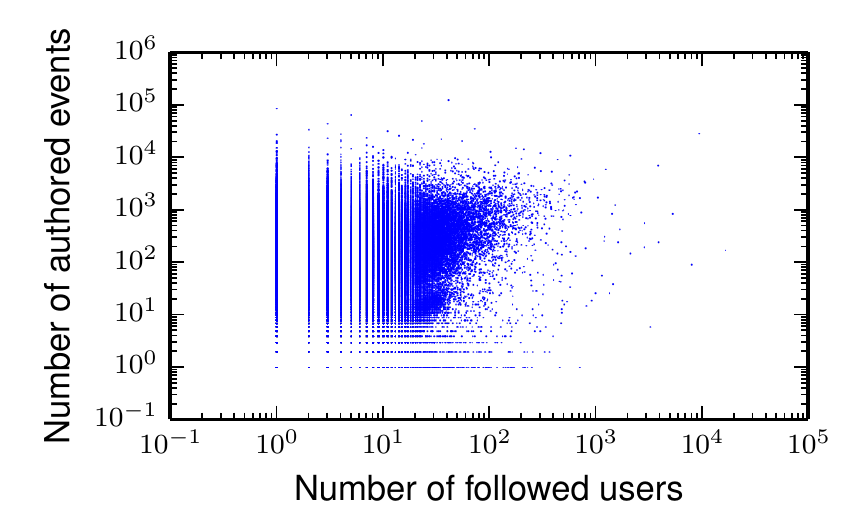}\label{fig:events-vs-following}}
\subfigure[Events vs Starred repositories]{\includegraphics[width=.45\linewidth]{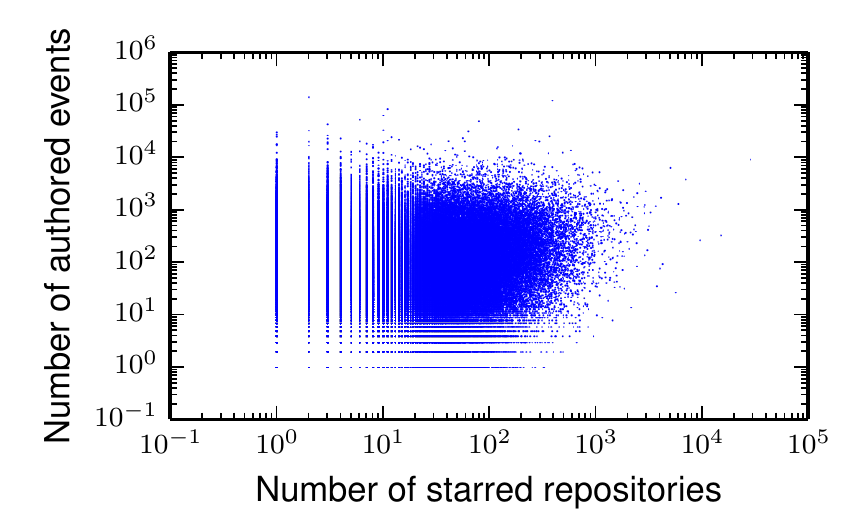}\label{fig:events-vs-starred}}
\caption{Number of actions executed by a user against (a) user followers, (b) number of repositories with write privileges, (c) followed users, (d) starred repositories.}
\end{figure*}

Human activities are commonly driven by reward mechanisms of some kind: people work to earn money and achieve a social status, they play games because they have fun, they travel because they enjoy seeing new places. A recent work has found that areas of brain connected to rewards are activated during the use of social networks websites~\cite{meshi_nucleus_2013}. One of the aspects that drives activity in GitHub, among others, is self-promotion~\cite{dabbish2012social}. We hypothesize that for a hybrid service like GitHub, both a social network and a collaboration network, some kind of indirect reward mechanism might and potentially underpin user activity. Even if it is not possible to provide definitive evidence about that, in the following we will show some interesting correlations between the activity of a user and some indirect rewards in terms of ``social prestige'' in GitHub.

In social networks, a common measure of user popularity and influence is given by the in-degree~\cite{wasserman_social_1994}. Therefore, it is reasonable to consider new connections as sort of rewards for users receiving them, as they increase their popularity. In order to investigate this aspect in GitHub we will search for correlation between user activity and degrees in the several graphs we have introduced. In Fig.~\ref{fig:events-vs-followers} we plot the number of authored events (i.e., for which the user \textit{actively} executes an action) for each user against the number of followers. We firstly note that people with a higher number of followers are commonly more active and people with lower levels of activity generally have fewer followers. However, we also observe that many users with a very high number of events have a very low number of followers: a higher level of activity does not directly translate into a larger number of followers. A similar phenomenon is also visible in Fig.~\ref{fig:events-vs-collaborators}, where we plot the number of authored events against the number of repositories for which a user is a collaborator or the repository owner. Being the collaborator can also be seen as a kind of indirect reward, as it is more important and prestigious than being a contributor. Collaborators receive permissions to modify the repository, whereas contributors only contribute their code through pull requests.

We are also interested to see whether a higher out-degree on the social graph is an indicator of a higher activity. However, in Fig.~\ref{fig:events-vs-following} it is possible to note that a much weaker correlation between these two quantities is present. A similar behavior can be observed in Fig.~\ref{fig:events-vs-starred}, where we plot activity versus the number of starred (i.e., bookmarked) repositories. In other words, users who follow many other users or bookmark many repositories are not much more active than those who do not.

\section{The Geography of Collaboration}
\begin{table}[t]
\center
    \scriptsize
    \begin{tabular}{ccc}
    \textbf{Rank} & \textbf{Country}        & \textbf{\%} \\ \hline
    1    & USA         & 30.14             \\
    2    & UK          & 6.43              \\
    3    & Germany     & 5.28              \\
    4    & China       & 5.11              \\
    5    & India       & 4.05              \\
    6    & France      & 3.87              \\
    7    & Canada      & 3.69              \\
    8    & Brazil      & 3.60              \\
    9    & Russia      & 3.14              \\
    10   & Japan       & 2.83              \\
    11   & Australia   & 2.00              \\
    12   & Spain       & 1.92              \\
    13   & Netherlands & 1.84              \\
    14   & Sweden      & 1.51              \\
    15   & Ukraine     & 1.37              \\
    16   & Italy       & 1.32              \\
    17   & Poland      & 1.02              \\
    18   & Switzerland & 0.86              \\
    19   & Belgium     & 0.75              \\
    20   & Mexico      & 0.74              \\
    \end{tabular}
    \quad \quad \quad
        \begin{tabular}{ccc}
 \textbf{City} & \textbf{\%} \\ \hline
    San Francisco, US & 3.84 \\
    London, GB        & 3.33 \\
    New York City, US & 2.93 \\
    Beijing, CN       & 1.98 \\
    Paris,  FR        & 1.80 \\
    Tokyo, JP         & 1.69 \\
    Seattle, US       & 1.59 \\
    Berlin, DE        & 1.49 \\
    Chicago, US       & 1.39 \\
    Shanghai, CN      & 1.34 \\
    Bangalore, IN     & 1.32 \\
    Toronto, CA       & 1.23 \\
    Moscow, RU        & 1.17 \\
    Austin, US    & 1.12 \\
    Boston, US    & 1.07 \\
    Los Angeles, US   & 1.01 \\
    Sydney, AU        & 0.94 \\
    Portland, US      & 0.88 \\
    Melbourne, AU     & 0.85 \\
    Stockholm, SE     & 0.81 \\
    \end{tabular}
    \caption{Top 20 countries and cities, ranked by absolute number of users.}
    \label{table:countries}
\end{table}

\begin{figure*}[tbh!]
\includegraphics[width=\linewidth]{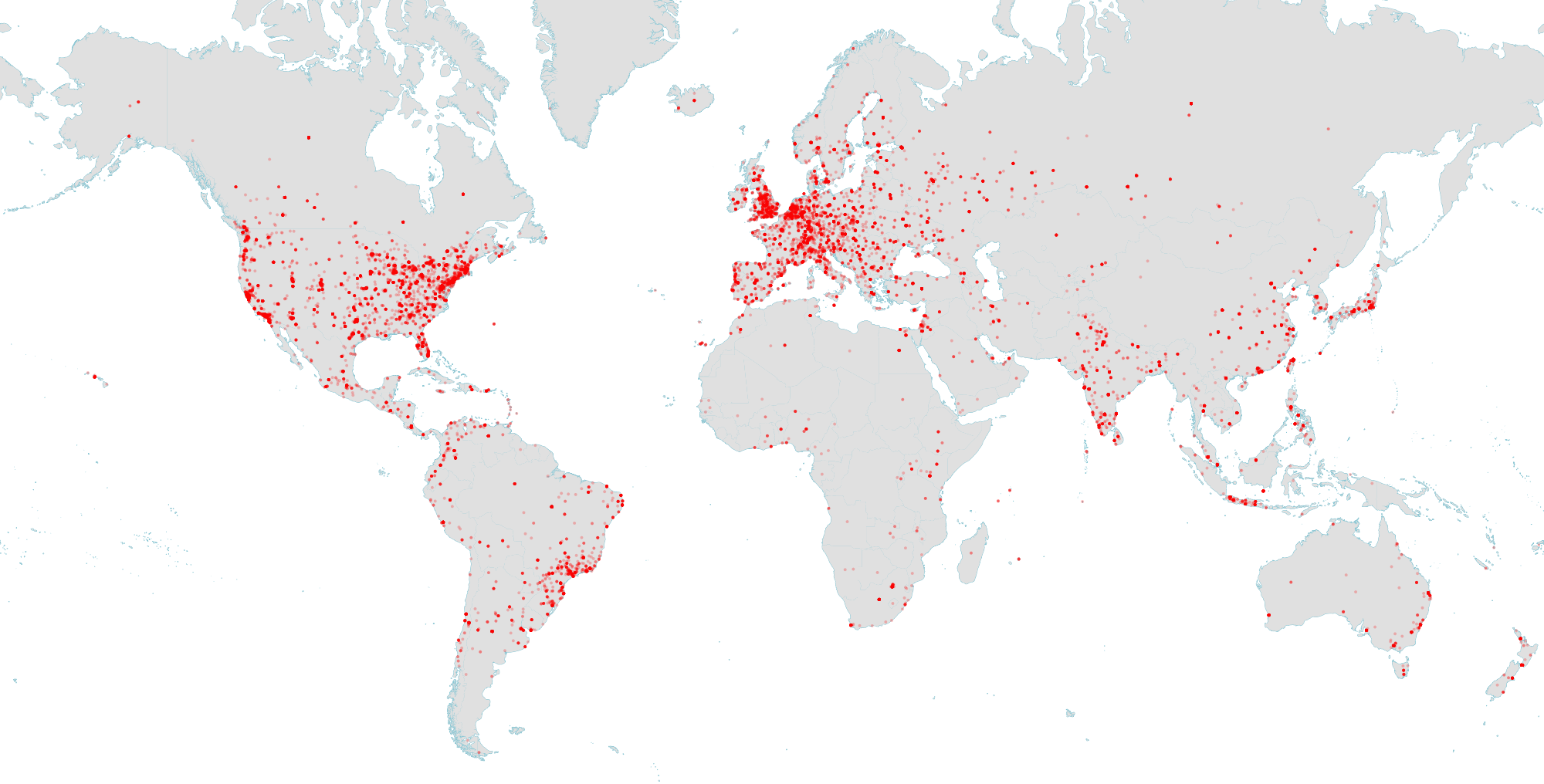}
    \caption{Distribution of GitHub users in the world. For each user, a partially transparent point is drawn on the map. The majority of users is located in North America and in Europe. The leading countries are the United States and the United Kingdom. A 15\% random sample of the original distribution was used to make this figure.}
    \label{fig:users-world}
\end{figure*}

In Fig.~\ref{fig:users-world} we show the geographic distribution of users in our dataset around the world. The majority of users is located in Europe and North America while other geographic regions have a consistently smaller number of users. The Tab.~\ref{table:countries}, listing the 20 most common countries and cities indicated in GitHub user profiles, confirm this consideration. The popularity of GitHub among developers living in the USA is really prominent, as 3 users out of 10 are based there.

%\subsection{Homophily}

%Several works in the literature suggest that in social networks interactions occur prevalently between similar people\mm{add references here}. In this section we will investigate whether that happens also on GitHub. More specifically, we explore how interactions between users can be explained as a result of them being in the same geographic area, being prolific (i.e., collaborating to many projects), having some favorite programming languages, or being similarly popular (in terms of number of followers).
%\mm{this part about homophily of programming languagesm, etc. is not presented - I removed this - it is more physical proximity}

\subsection{Impact of Geographic Proximity}
We also analyze the impact of physical proximity on the patterns of collaboration between different users.
%geographic similarity between connected users. 
Are people more likely to follow people who are closer to them? In Fig.~\ref{fig:follow-distance} we show the distribution of the distance covered by each pair of users connected by a directed link in the social network. The first part of the distribution shows that links decrease with distance, until $x=5000~km$: these are intra-continental links. The sudden drop at $x=5000~km$ is due to the ocean separating North America and Europe, that are the two regions where GitHub is mostly popular. For larger distances, the distribution increases again, showing a big presence of intercontinental links. This analysis, however, considers all the links, without discriminating them on a per-user basis.

\begin{figure}[th]
\includegraphics[width=\linewidth]{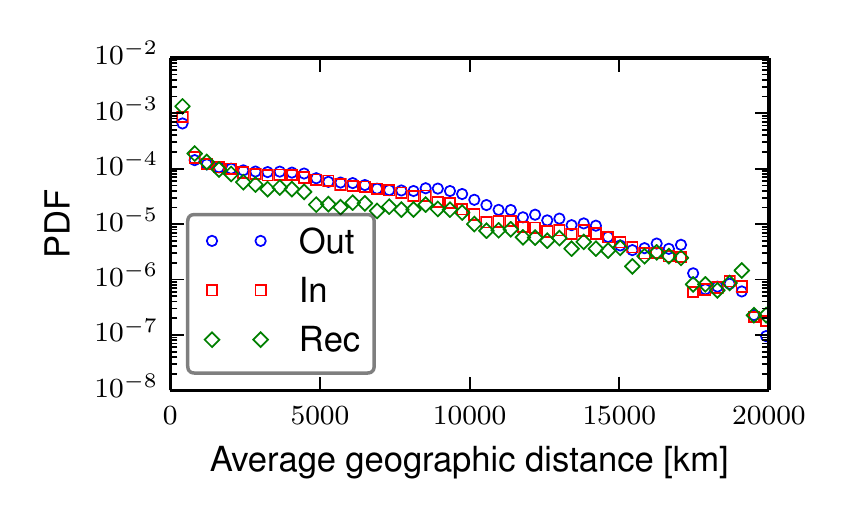}
    \caption{Distributions of the average geographic distance of a user's outgoing, incoming and reciprocated links, respectively represented by blue circles, red squares and green diamonds.}
\label{fig:avg-follow-distance}
\end{figure}

We now want to see how \textit{local} or \textit{global} is the neighborhood of a user, depending on how far his connections are located. In order to do that, we calculate for each user the average distance of their followers, their followed users and reciprocated links (i.e. users that are both followers and followed). In Fig.~\ref{fig:avg-follow-distance} we show the probability density function of the values of this measure. As expected, the distribution of these values decreases as the distance increases, indicating that users tend to interact with people that are close. We also notice that in the majority of the cases the average distances of the reciprocated connections of a user, usually considered as evidence of friendship or mutual acquaintance or collaboration, tend to be smaller compared to the other two types of links. %\al{How to explain the values at the end? Outliers?}\mm{potentially yes}

\subsection{Globality and Distant Collaboration}

%Similarly to what previously with follow relations,
%As we did for the follow relations, we would like to 
%
We now investigate if geographic proximity has an impact on the collaboration between users. In this case, we cannot compute the geographic distance between collaborators of a certain repository and the repository itself, as we cannot assign geographic coordinates to a repository. Project collaborators might be sparse around the globe or concentrated in a single city. 
In order to quantify how sparse they are, we define the \textit{globality} of a set of users $\mathcal{S}$ as follows:

\begin{figure}[th]
\includegraphics[width=\linewidth]{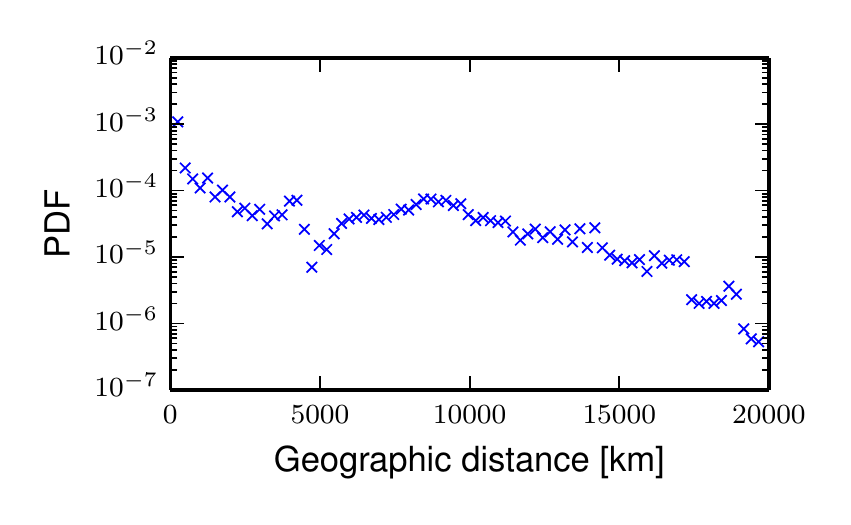}
    \caption{Distributions of the inter-users distance covered by each follow link. The distribution has a maximum at the lowest distance and gradually decreases for high distances.}
\label{fig:follow-distance}
\end{figure}

\begin{equation} G = \frac{1}{N d_{max}} \sum _{i,j \in \mathcal{S}}  d_{ij} \end{equation}
where $d_{max}$ is the maximum distance between two points on Earth where two generic users are localized and $N$ is the number of users taken into consideration. This measure is the normalized average of distances between all the points in the set. %It ranges between $0$, at which we have pure locality (i.e. all the users at the same point) and $1$, at which we have pure globality (i.e. users located at the antipodes). 
When all points coincide the measure is 0, whereas when the points are evenly distributed at the antipodes the measure is 1.

\begin{figure}[tb]
\includegraphics[width=\linewidth]{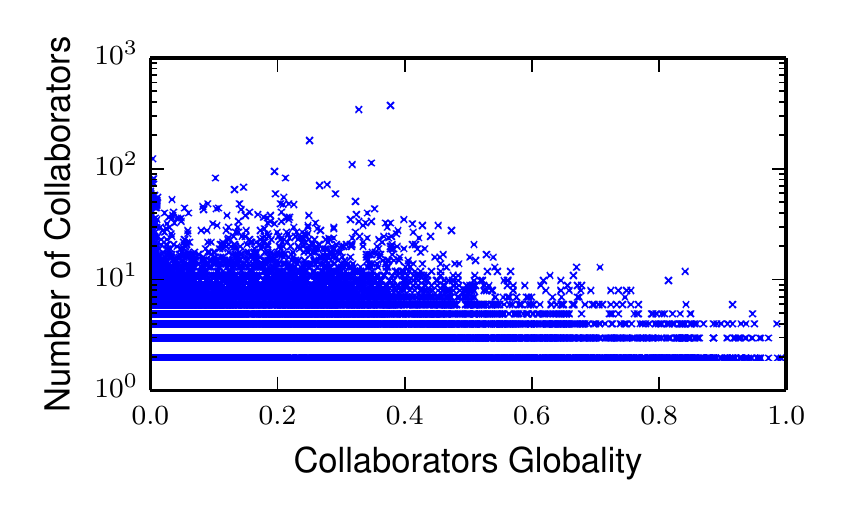}
    \caption{Scatter plot showing, for each repository, the number collaborators and the globality calculated over their geographic points. Intriguingly, repositories with a high number of collaborators exhibit smaller values of globality.}
        \label{fig:globality-collaboration}
\end{figure}

In Fig.~\ref{fig:globality-collaboration} we plot the value of globality against the number of collaboration for all the repositories that have at least two collaborators with location information in their profile.
Although points are quite dispersed in the plot we can make some considerations. For repository with a low number of collaborators, globality reaches values close to its maximum. As the number of collaborators goes up, the value of globality is found to be lower. This suggests that repositories with a low number of collaborators tend to have them concentrated around one or more key locations rather than scattered around the globe.

\begin{figure*}[t]
\centering
\subfigure[Followers]{\includegraphics[width=.38\linewidth]{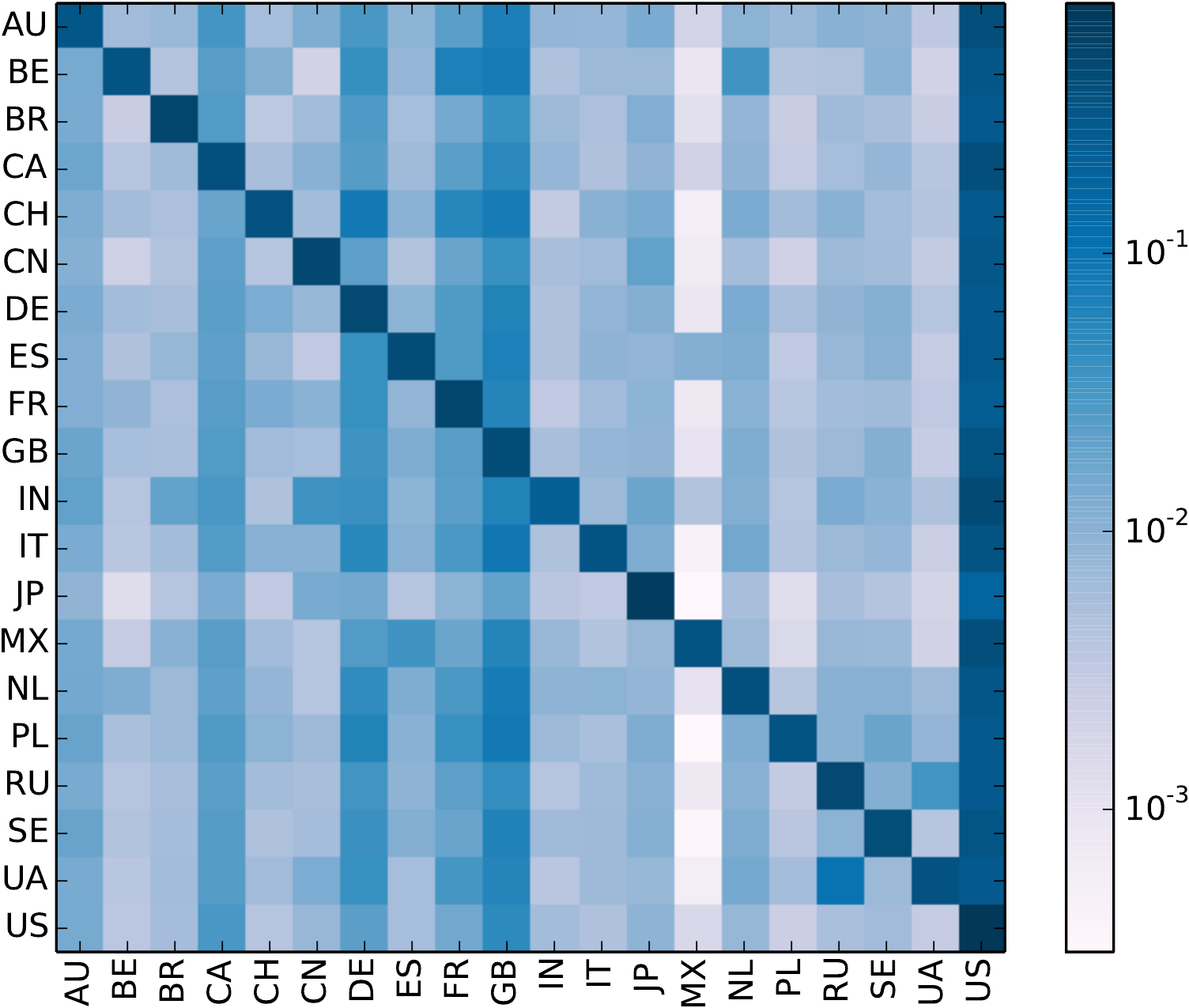}}
\hspace{0.2in}
\subfigure[Contributors]{\includegraphics[width=.38\linewidth]{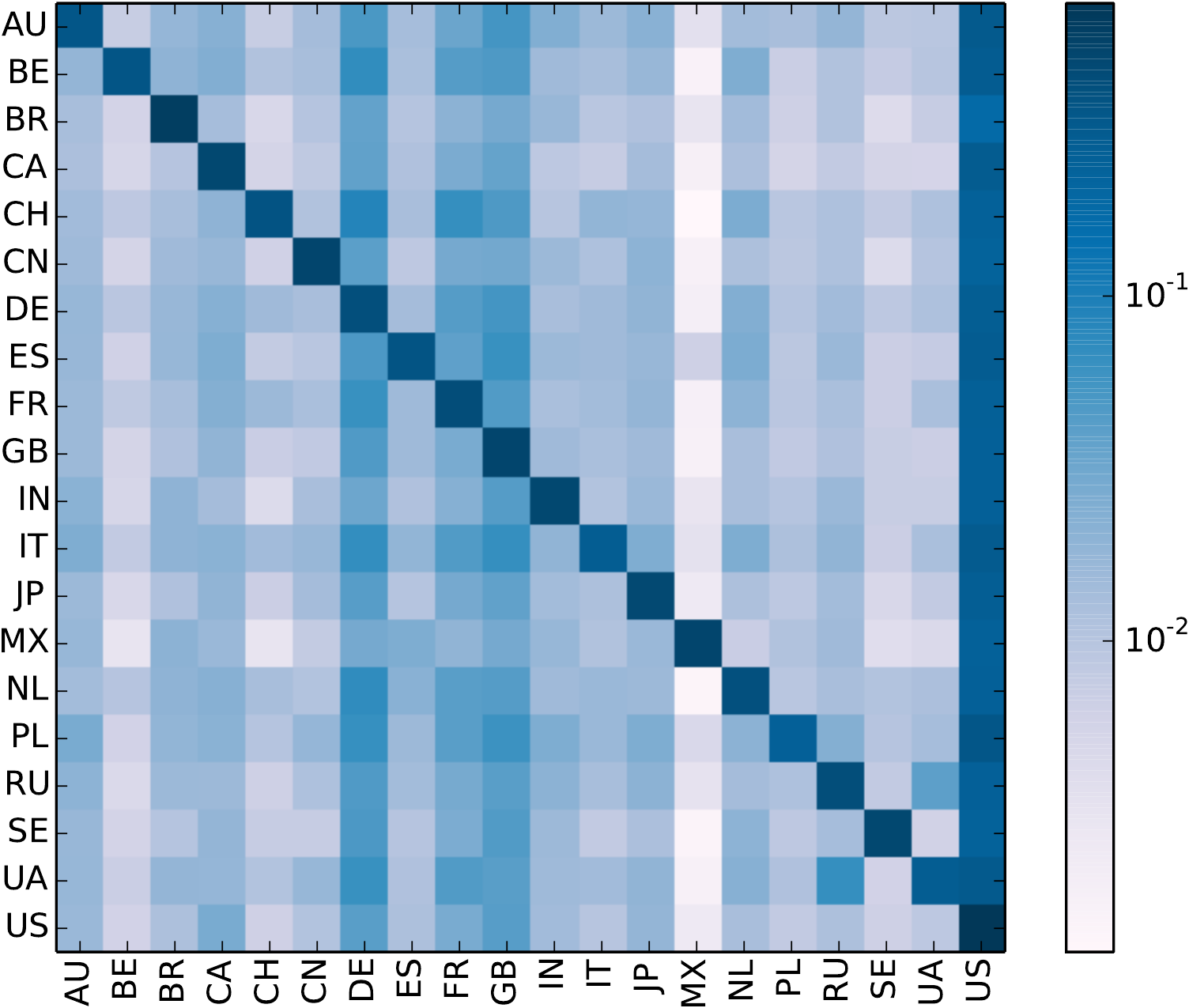}}
\caption{The ties among users of the top 20 countries in terms of number of users. The rows of the matrix are normalized to sum to unity. Note that the most followed users are in the United States, while the least followed country is Mexico. As expected, the matrix also shows a strong level of interaction between users from the same country. The left matrix is built from the followers graph, while the right matrix is built from the contributors graph.}
\label{fig:top-country}
\end{figure*}

We now investigate how social and collaborations links are distributed among countries. In order to do that, we build two square matrices $M_{G_F}$ and $M_{G_C^\perp}$ describing the number of links between countries. An element $m_{ij}$ of the matrices indicates the number of links from people in country $i$ to people in country $j$. The rows are normalized to sum to unity. This matrix has a precise meaning: each row shows how links coming from the people in a given country are distributed geographically. For clarity, in Fig.~\ref{fig:top-country} we show the normalized matrices only for the top 20 countries, although the following measures are calculated on the full matrices. We first note that both matrices have a strong diagonal component (on average $0.245$ and $0.346$ for followers and collaborators, respectively), in accordance with the fact that links are more likely to be directed to the same country of origin. The two matrices are also significantly similar, as confirmed by the low average cosine distance, amounting to $0.277$.

% flatex input end: [results.tex]

%
% flatex input: [conclusion.tex]
\section{Conclusions}
\label{conclusions}

In this paper we have analyzed the events happening on GitHub, the most popular repository for open source code, for 18 months between March 11, 2012 and September 11, 2013. We have obtained information about 2.19 million users and 5.68 million repositories. From this dataset we have derived four networks: a bipartite network describing the collaborations of users on repositories, a bipartite network describing the stars (bookmarks) assigned by users to repositories, a bipartite network describing the contributions of users on repositories and a directed social network describing the follow relations between users. We have showed that the distributions of the number of collaborators per project, contributors per project, stargazers per project and user followers show a power-law-like shape. We have found a very low reciprocity of the social ties, which is remarkably different from results of studies in other social networks; we have also observed that collaboration between users happens on a small fraction projects. We have found that very active users do not necessarily have a large number of followers. Finally, we have investigated the impact of geography on collaboration. Consistently to what happens in other social networks, users tend to interact with people that are close, as long-range links have a higher cost. A similar consideration can be made for repositories with a high number of collaborators, which tend to be managed by collaborators gravitating around specific locations.

We believe that our work provides novel insights about the complex dynamics of collaboration on a planetary scale. Our future research agenda includes the investigation of the software engineering issues that emerge from our quantitative analysis, especially with respect to the flow of information (and knowledge) that is present in the network of users. We think that this might represent a starting point for the development of novel strategies and tools for supporting online collaboration more effectively and efficiently. 
%\section*{Acknowledgements}
%This work was supported through the EPSRC Grant ``The Uncertainty of Identity: Linking Spatiotemporal Information Between Virtual and Real Worlds'' (EP/J005266/1).
%\section*{Acknowledgements}
%This work was supported through the EPSRC Grant ``The Uncertainty of Identity: Linking Spatiotemporal Information Between Virtual and Real Worlds'' (EP/J005266/1).
% flatex input end: [conclusion.tex]

%

\section*{Acknowledgments}
This work was supported through the EPSRC Grant ``The Uncertainty of Identity: Linking Spatiotemporal Information Between Virtual and Real Worlds'' (EP/J005266/1).

%FLATEX-REM:\bibliography{github}

\begin{thebibliography}{}

\bibitem[\protect\citeauthoryear{Bird \bgroup et al\mbox.\egroup
  }{2008}]{bird2008latent}
Bird, C.; Pattison, D.; D'Souza, R.; Filkov, V.; and Devanbu, P.
\newblock 2008.
\newblock Latent social structure in open source projects.
\newblock In {\em Proceedings of FSE'08},  24--35.
\newblock ACM.

\bibitem[\protect\citeauthoryear{Brandes \bgroup et al\mbox.\egroup
  }{2009}]{brandes2009network}
Brandes, U.; Kenis, P.; Lerner, J.; and van Raaij, D.
\newblock 2009.
\newblock {Network Analysis of Collaboration Structure in Wikipedia}.
\newblock In {\em Proceedings of WWW'09},  731--740.
\newblock ACM.

\bibitem[\protect\citeauthoryear{Cha, Mislove, and
  Gummadi}{2009}]{cha2009measurement}
Cha, M.; Mislove, A.; and Gummadi, K.~P.
\newblock 2009.
\newblock A measurement-driven analysis of information propagation in the
  flickr social network.
\newblock In {\em Proceedings of WWW'09},  721--730.
\newblock ACM.

\bibitem[\protect\citeauthoryear{Clauset, Shalizi, and
  Newman}{2009}]{clauset2009power}
Clauset, A.; Shalizi, C.~R.; and Newman, M.~E.
\newblock 2009.
\newblock Power-law distributions in empirical data.
\newblock {\em SIAM review} 51(4):661--703.

\bibitem[\protect\citeauthoryear{Colizza \bgroup et al\mbox.\egroup
  }{2006}]{colizza2006detecting}
Colizza, V.; Flammini, A.; Serrano, M.~A.; and Vespignani, A.
\newblock 2006.
\newblock Detecting rich-club ordering in complex networks.
\newblock {\em Nature Physics} 2(2):110--115.

\bibitem[\protect\citeauthoryear{Dabbish \bgroup et al\mbox.\egroup
  }{2012}]{dabbish2012social}
Dabbish, L.; Stuart, C.; Tsay, J.; and Herbsleb, J.
\newblock 2012.
\newblock {Social Coding in GitHub: transparency and collaboration in an open
  software repository}.
\newblock In {\em Proceedings of CSCW'12},  1277--1286.
\newblock ACM.

\bibitem[\protect\citeauthoryear{Finley}{2011}]{finley_github_2011}
Finley, K.
\newblock 2011.
\newblock {GitHub} has surpassed {Sourceforge} and {Google Code} in popularity.
\newblock \url{http://readwrite.com/2011/06/02/github-has-passed-sourceforge}.

\bibitem[\protect\citeauthoryear{Gousios and
  Spinellis}{2012}]{gousios2012ghtorrent}
Gousios, G., and Spinellis, D.
\newblock 2012.
\newblock {GHTorrent}: Github's data from a firehose.
\newblock In {\em Proceedings of {MSR}'12},  12--21.

\bibitem[\protect\citeauthoryear{Heller \bgroup et al\mbox.\egroup
  }{2011}]{heller2011visualizing}
Heller, B.; Marschner, E.; Rosenfeld, E.; and Heer, J.
\newblock 2011.
\newblock Visualizing collaboration and influence in the open-source software
  community.
\newblock In {\em Proceedings of {MSR}'11},  223--226.

\bibitem[\protect\citeauthoryear{Hindle, German, and
  Holt}{2008}]{hindle_what_2008}
Hindle, A.; German, D.~M.; and Holt, R.
\newblock 2008.
\newblock What do large commits tell us?: a taxonomical study of large commits.
\newblock In {\em Proceedings of {MSR}'08}.
\newblock New York, {NY}, {USA}: {ACM}.

\bibitem[\protect\citeauthoryear{Kittur \bgroup et al\mbox.\egroup
  }{2007}]{kittur_power_2007}
Kittur, A.; Chi, E.; Pendleton, B.; Suh, B.; and Mytkowicz, T.
\newblock 2007.
\newblock Power of the few vs. wisdom of the crowd: Wikipedia and the rise of
  the bourgeoisie.
\newblock {\em Proceedings of {CHI}'07}.

\bibitem[\protect\citeauthoryear{Kumar, Novak, and
  Tomkins}{2010}]{kumar2010structure}
Kumar, R.; Novak, J.; and Tomkins, A.
\newblock 2010.
\newblock Structure and evolution of online social networks.
\newblock In {\em Link Mining: Models, Algorithms, and Applications}. Springer.
\newblock  337--357.

\bibitem[\protect\citeauthoryear{Kwak \bgroup et al\mbox.\egroup
  }{2010}]{kwak2010twitter}
Kwak, H.; Lee, C.; Park, H.; and Moon, S.
\newblock 2010.
\newblock What is twitter, a social network or a news media?
\newblock In {\em Proceedings of WWW'10},  591--600.
\newblock ACM.

\bibitem[\protect\citeauthoryear{Lehmann}{2006}]{lehmann2006nonparametrics}
Lehmann, E.
\newblock 2006.
\newblock {\em Nonparametrics: Statistical Methods based on Ranks (POD)}.
\newblock Prentice-Hall.

\bibitem[\protect\citeauthoryear{Lunden}{2013}]{techcrunch_github_2013}
Lunden, I.
\newblock 2013.
\newblock {GitHub} hits the {4M} user mark as it looks beyond developers for
  its next stage of growth.
\newblock
  \url{http://techcrunch.com/2013/09/11/github-hits-the-4m-user-mark-as-it-looks-beyond-developers-for-its-next-stage-of-growth/}.

\bibitem[\protect\citeauthoryear{Meshi, Morawetz, and
  Heekeren}{2013}]{meshi_nucleus_2013}
Meshi, D.; Morawetz, C.; and Heekeren, H.~R.
\newblock 2013.
\newblock Nucleus accumbens response to gains in reputation for the self
  relative to gains for others predicts social media use.
\newblock {\em Frontiers in Human Neuroscience} 7:439.

\bibitem[\protect\citeauthoryear{Newman}{2002}]{newman2002assortative}
Newman, M.~E.
\newblock 2002.
\newblock Assortative mixing in networks.
\newblock {\em Physical Review Letters} 89(20):208701.

\bibitem[\protect\citeauthoryear{Shrestha, Zhu, and
  Miller}{2013}]{shrestha2013visualizing}
Shrestha, A.; Zhu, Y.; and Miller, B.
\newblock 2013.
\newblock Visualizing time and geography of open source software with
  storygraph.
\newblock In {\em Proceedings of VISSOFT'13},  1--4.
\newblock IEEE.

\bibitem[\protect\citeauthoryear{Valverde and
  Sol{\'e}}{2007}]{valverde2007self}
Valverde, S., and Sol{\'e}, R.~V.
\newblock 2007.
\newblock Self-organization versus hierarchy in open-source social networks.
\newblock {\em Physical Review E} 76(4):046118.

\bibitem[\protect\citeauthoryear{Vuong \bgroup et al\mbox.\egroup
  }{2008}]{vuong_ranking_2008}
Vuong, B.-Q.; Lim, E.-P.; Sun, A.; Le, M.-T.; Lauw, H.~W.; and Chang, K.
\newblock 2008.
\newblock On ranking controversies in wikipedia: Models and evaluation.
\newblock In {\em Proceedings of {WSDM}'08},  171–182.
\newblock New York, {NY}, {USA}: {ACM}.

\bibitem[\protect\citeauthoryear{Wasserman and
  Faust}{1994}]{wasserman_social_1994}
Wasserman, S., and Faust, K.
\newblock 1994.
\newblock {\em Social Network Analysis: Methods and Applications}.
\newblock Cambridge University Press, 1 edition.

\bibitem[\protect\citeauthoryear{Watts and
  Strogatz}{1998}]{watts1998collective}
Watts, D.~J., and Strogatz, S.~H.
\newblock 1998.
\newblock Collective dynamics of small-world networks.
\newblock {\em Nature} 393(6684):440--442.

\bibitem[\protect\citeauthoryear{Zhou and Mondrag{\'o}n}{2004}]{zhou2004rich}
Zhou, S., and Mondrag{\'o}n, R.~J.
\newblock 2004.
\newblock The rich-club phenomenon in the internet topology.
\newblock {\em IEEE Communications Letters} 8(3):180--182.

\end{thebibliography}
%*flatex input: [github.bbl]

% flatex input end: [github.bbl]
%FLATEX-REM:\bibliographystyle{aaai}

\end{document}